\documentclass[onecolumn]{revtex4}
\usepackage{amssymb}
\usepackage{amsmath}
\usepackage{dcolumn}
\usepackage{bm}
\usepackage{graphicx}

\begin{document}

\title{Geometrical bounds on irreversibility in squeezed thermal bath}
\author{Chen-Juan Zou$^{1}$}
\author{Yue Li$^{2}$}
\author{Jia-Bin You$^{3}$}
\author{Qiong Chen$^{1}$}
\author{Wan-Li Yang$^{2}$}
\author{Mang Feng$^{2}$}
\affiliation{$^{1}$Key Laboratory of Low-Dimensional Quantum Structures and Quantum
Control of Ministry of Education, Key Laboratory for Matter Microstructure
and Function of Hunan Province, Department of Physics, Hunan Normal
University, Changsha 410081, China}
\affiliation{$^{2}$State Key Laboratory of Magnetic Resonance and Atomic and Molecular
Physics, Innovation Academy for Precision Measurement Science and
Technology, Chinese Academy of Sciences, Wuhan 430071, China}
\affiliation{$^{3}$Institute of High Performance Computing, Agency for Science, Technology, and Research (A*STAR), 1 Fusionopolis Way, \#16-16 Connexis, Singapore 138632.}

\begin{abstract}
Irreversible entropy production (IEP) plays an important role in quantum
thermodynamic processes. Here we investigate the geometrical bounds of IEP
in nonequilibrium thermodynamics by exemplifying a system coupled to a
squeezed thermal bath subject to dissipation and dephasing, respectively. We
find that the geometrical bounds of the IEP always shift in contrary way
under dissipation and dephasing, where the lower and upper bounds turning to
be tighter occurs in the situation of dephasing and dissipation,
respectively. However, either under dissipation or under dephasing, we may
reduce both the critical time of the IEP itself and the critical time of the
bounds for reaching an equilibrium by harvesting the benefits of squeezing
effects, in which the values of the IEP, quantifying the degree of
thermodynamic irreversibility, also becomes smaller. Therefore, due to the
nonequilibrium nature of the squeezed thermal bath, the system-bath
interaction energy brings prominent impact on the IEP, leading to tightness
of its bounds. Our results are not contradictory with the second law of
thermodynamics by involving squeezing of the bath as an available resource,
which can improve the performance of quantum thermodynamic devices.

\textbf{Keywords:}  squeezed thermal bath, irreversible entropy production,
geometrical bounds
\end{abstract}

\maketitle

\section{Introduction}

Over the past two decades, the nonequilibrium phenomena and\emph{\ }%
thermodynamic irreversibility quantified by irreversible entropy production
(IEP) have drawn much attention, since this fundamental concept is one of
the cornerstones of classical and quantum thermodynamics \cite%
{TM1,TM2,TM3,TM4,TM5,TM6,TM7,TM8,TM9,TM10}. As is well known, the positivity
of entropy production has been universally captured by the conventional
second law of thermodynamics (SLT) \cite{TM9}, which quantitatively
characterize the interplay between the exchange of energy and the
irreversibility by introducing the state function entropy \cite{fun1,fun2}.
Also, through specifying a lower bound for the irreversible entropy change,
the related Clausius inequality provides a fundamental feature of
irreversible phenomena. Noted that this lower bound (zero) is trivially
independent of how far from equilibrium a process operates \cite{TM6}.

Recently, with restriction to a specific class of nonequilibrium phenomena,
such as thermal relaxation process, rich features of thermodynamic
irreversibility have been found successively \cite{TM2}. Such a lower bound
for classical, near-equilibrium transformation processes has been derived by
mean of a geometric approach \cite{rmp1}. S. Deffner and E. Lutz treated the
system's Hilbert space as a Riemannian manifold and\textbf{\ }extended the
classical case \cite{rmp1} to the nonequilibrium closed quantum system. The
obtained generalized Clausius inequality states that the thermodynamic
irreversibility is bounded in terms of the Bures length between the final
state and the corresponding equilibrium state, by using information geometry
\cite{TM6}\emph{. }Soon afterwards, they broadened the closed quantum system
further\emph{\ }to the weakly coupled open quantum system, then obtained the
exact microscopic expressions for the nonequilibrium entropy production \cite%
{TM7}. Along this direction, reference \cite{fun1} theoretically and
experimentally determined a sharper geometrical bound for a qubit
thermalization process and obtained a tighter version of the Clausius
inequality following a similar approach.

In many cases of interest, however, having a sharper lower bound is
essential. Therefore, theses above-mentioned progress hence stimulated
successive studies on the tightness of the geometrical bounds on
irreversibility in open quantum system \cite{TM6,TM7,TM8,TM9,TM10,fun1}.
These recent publications tried to develop theories to further understand
the thermodynamic irreversibility inherent to nonequilibrium processes. In
particular, based on the variational principle and time-reversed map, the
authors in reference \cite{TM10} obtained a information-theoretical bound
for entropy production in a relaxation process by a geometrical distance on
the Riemannian manifold \cite{Rie}, which was experimentally validated by a
single ultracold trapped ion $^{40}$Ca \cite{Ca}.

T. V. Vu and Y. Hasegawa strengthened the Clausius inequality and proved
that IEP is bounded from below by a modified Wasserstein distance (quantum
generalization of the Wasserstein metric) between the initial and final
states \cite{TM8}.\emph{\ }Thereafter, they extended this single-bath case
to the case of multiple-bath, and refined the SLT in a quantum regime,
through deriving the fundamental\emph{\ }bound on irreversibility for
thermal relaxation processes of Markovian open quantum systems \cite{TM9}.
\emph{\ }

On the other hand, quantum bath engineering techniques are powerful tools
that enable the realization of arbitrary thermal and nonthermal environments
\cite{no1,no2,no3,no4,no5,no6,no7,no8}. Additionally, due to the unique
vantage of the quantum control, bath engineering has aroused widespread
interest in the context of quantum thermodynamical processes. Various
strategies have attempted to improve the the performance of thermodynamic
devices \cite{dev1,dev2,dev3,dev5,dev6,dev7,dev8,dev9,dev10}, whose
efficiency is usually reduced by the presence of IEP\emph{\ }\cite{TM7}. For
instance, the use of a squeezed thermal bath allows us to operate
thermodynamic devices beyond the classical bound \cite{no8,dev1}. In
particular, the experiment in reference \cite{dev8} demonstrated that the
efficiency of the quantum heat engine may go beyond the standard Carnot
efficiency by employing a squeezed thermal bath.
\begin{figure}[tbp]
\centering\includegraphics[width=7.5 cm,bb=5pt 1pt 293pt 215pt]{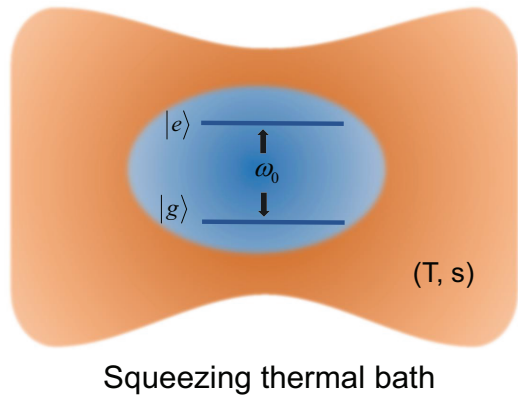}
\caption{Schematic diagram for two-level system (excited state $\left\vert
e\right\rangle $ and ground state $\left\vert g\right\rangle $) interacting
with a squeezed thermal bath at temperature $T$ with squeezing parameter $s$%
. The transition frequency between the two levels is $\protect\omega _{0}$.
The evolution induced by the system-bath interaction produces irreversible
entropy.}
\label{setup}
\end{figure}

However, the influence of squeezed thermal bath on the bounds of IEP in the
quantum thermodynamics has been largely unexplored. A precise
characterization of the IEP in such an unchartered domain, and a general
framework providing a deeper understanding of the associated quantum
thermodynamic phenomena therefore appear necessary. Therefore, it is
instructive to look into the role of squeezed thermal bath playing during
the process of thermodynamic irreversibility.

In this work, we study and quantify the geometrical bounds on
irreversibility of a quantum system contacting with a squeezed thermal bath.
Here we consider two distinct situations including the dissipation model and
dephasing model, respectively. Starting with Born-Markovian quantum master
equation, we derived analytical expressions for time-dependent reduced
density matrix of system, in which the squeezing parameter of bath is
involved. Then through quantifying the degree of irreversibility by the IEP,
we find that the geometrical bounds of IEP decrease (increase) with the
growth of degree of squeezing in the case of dissipation (dephasing) model,
respectively. Additionally, the common feature is that the critical times of
the IEP itself and that of IEP's bounds reaching equilibrium, as well as the
values of IEP quantifying the degree of thermodynamic irreversibility are
reduced due to the presence of squeezing effect, for these two models.
Therefore, due to the nonequilibrium nature of squeezed thermal bath, the
interaction energy between the system and bath brought important impact on
the irreversibility, as well as the tightness of its bounds. As expected,
our finding obeys the principles of thermodynamics and reveals richer
features of thermodynamics relaxation process, and the presence of a quantum
property, such as the squeezing effects included in the bath, could be used
to serve as an available resource to improve the performance of the quantum
thermodynamic devices.

The paper is organized as follows. We give a brief account of the method
about geometrical bounds on irreversibility in open quantum system (Section
II). We evaluate the geometrical bounds on irreversibility for squeezed
thermal bath in the dissipation model (Section III) and dephasing model
(Section IV), respectively, and we conclude and give prospects for future
developments in Section V.

\section{Geometrical bounds of irreversible entropy production}

Considering an arbitrary quantum system with the Hamiltonian coupled to a
thermal bath. Usually, the quantum system is initialized in a given state $%
\rho _{0}$, then interacts with a bath at temperature $T$. The evolution
induced by the interactions brings the system in a state $\rho (t)$ and
produces irreversible entropy. Furthermore, the system will thermalize with
the bath and then asymptotically reaching the unique canonical equilibrium
state $\rho _{th}$ if the system Hamiltonian $H$ maintains constant. The
total entropy variation of the system is defined as
\begin{equation}
\Delta S_{tot}=\Delta S_{ir}+\Delta S_{re}=S\left[ \rho (t)\right] -S\left[
\rho _{0}\right] ,  \label{entropy1}
\end{equation}%
where $S\left[ \rho \right] =-tr\left( \rho \ln \rho \right) $ is the von
Neumann entropy. The IEP (irreversible part of the total entropy variation $%
\Delta S_{tot}$) is denoted by

\begin{equation}
\Delta S_{ir}(t)=S\left( \rho _{0}\Vert \rho _{th}\right) -S\left( \rho
(t)\Vert \rho _{th}\right) ,  \label{entrop2}
\end{equation}%
which is the thermodynamic irreversibility under consideration in the
present work \cite{TM2,fun1}. $S\left( \rho _{1}\Vert \rho _{2}\right)
=tr\left( \rho _{1}\ln \rho _{1}\right) -tr\left( \rho _{1}\ln \rho
_{2}\right) $ represents the quantum relative entropy of $\rho _{1}$ to $%
\rho _{2}$. The entropy flow between the system and the environment
(reversible part of the total entropy variation $\Delta S_{tot}$) is $\Delta
S_{re}=\Delta Q/T$ , where $\Delta Q=tr\left( H\rho (t)\right) -tr\left(
H\rho _{0}\right) $ is the heat absorbed by the system \cite{fun1}. On the
other hand, the Clausius inequality $\Delta S_{ir}\geq 0$ putting forward
the lower limit of IEP is always non-negative. In order to deepen the
understanding of how much energy in the irreversible process is consumed, it
is essential to search a sharper or tighter bound on irreversibility.\emph{\
}By treating the Hilbert space of the system as a Riemannian manifold, the
relationship between IEP and the geodesic distance $D$ corresponding to the
metric that is contractive under complete positive and trace preserving maps
can be directly established, and a generalized form of Clausius inequality
can be obtained by deriving the Wigner-Yanase length between the initial and
final states of the system. Note that the only cases in which an analytical
expression for the geodesic distance is known are the Wigner-Yanase metric $%
D_{WY}\left( \rho _{1},\rho _{2}\right) =\arccos \left[ tr\left\{ \sqrt{\rho
_{1}}\sqrt{\rho _{2}}\right\} \right] $ and the quantum Fisher information
metric $D_{QF}\left( \rho _{1},\rho _{2}\right) =\arccos \left[ tr\left\{
\sqrt{\sqrt{\rho _{1}}\rho _{2}\sqrt{\rho _{1}}}\right\} \right] $\ \cite%
{fun1}. Based on these analytical expressions, one can obtain the\emph{\ }%
geometrical LB of IEP as

\begin{equation}
\Delta S_{ir}\left( t\right) \geq \frac{8}{\pi ^{2}}\max_{\left\{
X=QF,WY\right\} }D_{X}^{2}\left( \rho _{0},\rho (t)\right) ,  \label{lower}
\end{equation}%
and the\emph{\ }geometrical UB is

\begin{equation}
\Delta S_{ir}\left( t\right) \leq S\left( \rho _{0}\Vert \rho _{th}\right) -%
\frac{8}{\pi ^{2}}\max_{\left\{ X=QF,WY\right\} }D_{X}^{2}\left( \rho
(t),\rho _{th}\right) .  \label{upper}
\end{equation}

\begin{figure}[tbp]
\centering\includegraphics[width=11cm,bb=41pt 28pt 860pt 415pt]{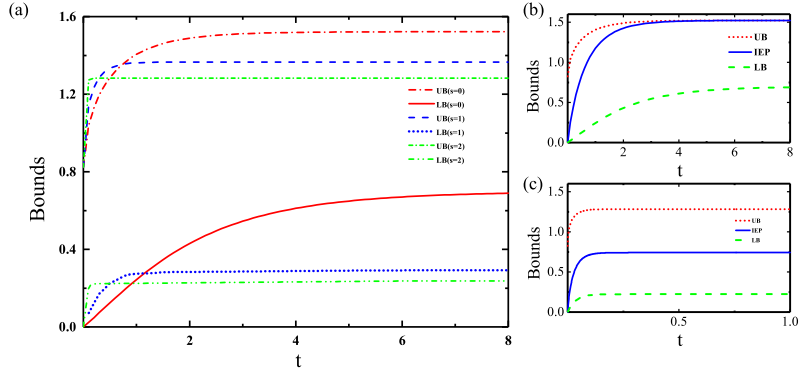}
\caption{(a) Time dependence of LB and UB of IEP under different degree of
squeezing $s$ in the dissipation model. Time dependence of IEP (blue solid
line) and its LB (red dotted line) and UB (green dashed line) in the case of
(b) $s=0$ and (c) $s=2$. The initial state of system is $\left( \left\vert
e\right\rangle +\left\vert g\right\rangle \right) /\protect\sqrt{2}$.
Hereafter, we choose $\protect\omega _{0}$ as the frequency unit, $T=0.34$,
and $\protect\phi =0.$}
\end{figure}

From the above two relations one can define the related bound gap as

\begin{equation}
\Delta U=UB-LB,  \label{DU}
\end{equation}%
and the deviation of the IEP from the LB or UB is given by
\begin{equation}
\Delta \delta _{L}=\Delta S_{ir}\left( t\right) -LB,  \label{deltaL}
\end{equation}%
or
\begin{equation}
\Delta \delta _{U}=UB-\Delta S_{ir}\left( t\right),  \label{deltaU}
\end{equation}%
respectively. In this paper, we say a LB (UB) is relative tighter if the LB
(UB) takes a larger (smaller) value compared with the case of conventional
thermal bath.

\section{Geometrical bounds on irreversibility in the dissipation model}

Here we consider the dissipation model taking into account the effect of the
squeezed thermal bath with temperature $T$, in the case of
single-excitation. The total Hamiltonian is (in units of $\hbar =1$)

\begin{eqnarray}
H &=&H_{S}+H_{B}+H_{SB}  \notag \\
&=&\omega _{0}\hat{\sigma}_{+}\hat{\sigma}_{-}+\sum\limits_{k}\omega _{k}%
\hat{b}_{k}^{\dagger }\hat{b}_{k}+\sum\limits_{k}(g_{k}\hat{\sigma}_{+}\hat{b%
}_{k}+H.c.),  \label{H1}
\end{eqnarray}%
where $H_{S}$, $H_{B}$, and $H_{SB}$ stand for the Hamiltonians of the
system, bath, and system-bath interaction, respectively. $\hat{\sigma}_{+}(%
\hat{\sigma}_{-})=\left\vert e\right\rangle \left\langle g\right\vert
(\left\vert g\right\rangle \left\langle e\right\vert )$ and $\omega _{0}$
are the inversion operator and transition frequency of the system with $%
\left\vert e\right\rangle $ and $\left\vert g\right\rangle $ being the
excited and ground states. $\hat{b}_{k}^{\dagger }(\hat{b}_{k})$ are the
creation (annihilation) operators of the k-th mode of the bath. The coupling
strength between the system and the bath is denoted by $g_{k}$.

The master equation, in the interaction picture, is given by the following
Lindblad form \cite{master1,master2}%
\begin{eqnarray}
\dot{\rho}_{s}(t) &=&\frac{\gamma N}{2}D[\sigma _{+}]\rho _{s}(t)+\frac{%
\gamma \left( 1+N\right) }{2}[\sigma _{-}]\rho _{s}(t)  \notag \\
&&-\gamma M\sigma _{+}\rho _{s}(t)\sigma _{+}-\gamma M^{\ast }\sigma
_{-}\rho _{s}(t)\sigma _{-},  \label{M1}
\end{eqnarray}%
where $\dot{\rho}_{s}(t)=d\rho _{s}(t)/dt$, and $D[A]\rho =2A\rho
A^{+}-A^{+}A\rho -\rho A^{+}A$. The spontaneous emission\emph{\ }rate of
system is $\gamma $ and $N=N_{th}\left[ \cosh ^{2}\left( s\right) +\sinh
^{2}\left( s\right) \right] +\sinh ^{2}\left( s\right) $, and $M=-\sinh
\left( 2s\right) e^{i\phi }\left( 2N_{th}+1\right) /2$ with $s$ and $\phi $
the bath squeezing strength and phase, respectively. $N_{th}=1/\left(
e^{\omega _{0}\beta }-1\right) $ is the Plank distribution, where $\beta
=1/k_{B}T$ with $k_{B}=1$ the Boltzmann constant.

\begin{figure*}[tbp]
\centering\includegraphics[width=14cm,bb=2pt 6pt 809pt 343pt]{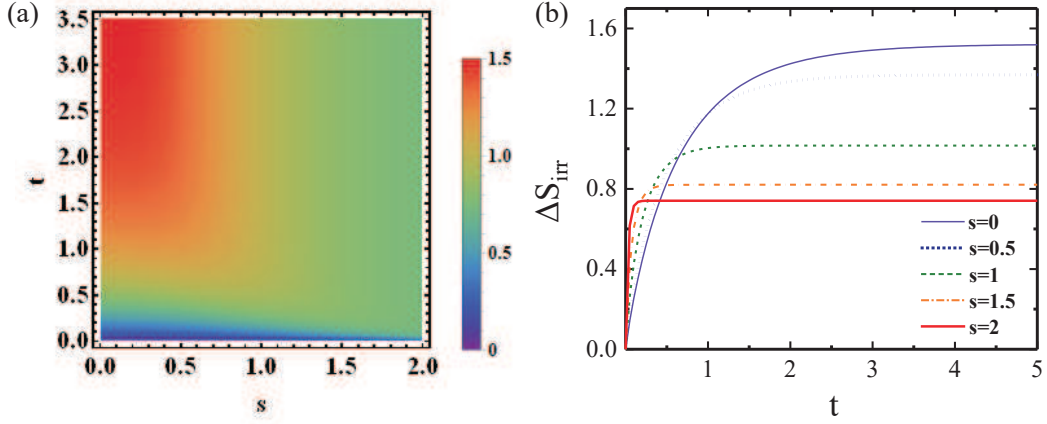}
\caption{(a) The values of the IEP $\triangle S_{irr}$ in the parameter
plane of $\{s,t\}$ in the dissipation model. (b) Time dependence of IEP $%
\triangle S_{irr}$ under the different degrees of squeezing $s$. }
\end{figure*}

Rewriting the density matrix as $\rho _{s}(t)$ $=\left( \mathbf{I}+\vec{r}%
\left( t\right) \cdot \vec{\sigma}\right) /2$ by mean of the Bloch vector $%
\vec{r}\left( t\right) =Tr[\vec{\sigma}\rho _{s}(t)]$ with the Identity
matrix $\mathbf{I}$, we can transform the master equation (\ref{M1}) into
the Bloch equation%
\begin{equation}
\frac{d}{dt}\vec{r}\left( t\right) =\xi \vec{r}\left( t\right) +\vec{m},
\label{Bloch1}
\end{equation}%
with%
\begin{equation}
\xi =\left(
\begin{array}{ccc}
-\frac{\tilde{\gamma}+2\gamma M}{2} & 0 & 0 \\
0 & -\frac{\tilde{\gamma}-2\gamma M}{2} & 0 \\
0 & 0 & -\tilde{\gamma}%
\end{array}%
\right)  \label{Bloch2}
\end{equation}%
and $\vec{m}=(0,0,,-\gamma _{0})^{T}$. Here $\tilde{\gamma}=\gamma \left(
2N+1\right) $ is the total transition rate. Assuming the system to be
initially in the ground state $\left\vert g\right\rangle $, a
straightforward calculation yields the analytical solution

\begin{equation}
\rho _{s}(t)=\left(
\begin{array}{cc}
\frac{1-\nu +(1-\nu )\left\langle \sigma _{z}\right\rangle _{ss}}{2} & \mu
\left\langle \sigma _{-}\right\rangle _{ss} \\
(1-e^{\frac{-\gamma _{s}t}{2}})\left\langle \sigma _{+}\right\rangle _{ss} &
\frac{1+\nu -(1-\nu )\left\langle \sigma _{z}\right\rangle _{ss}}{2}%
\end{array}%
\right) ,  \label{sol}
\end{equation}%
Here $\left\langle \sigma _{\pm }\right\rangle _{ss}$ and $\left\langle
\sigma _{z}\right\rangle _{ss}\ $are the stationary solutions of
differential equation (\ref{Bloch1}). $\mu =\{\gamma _{s}-e^{\left( -4\tilde{%
\gamma}+\gamma _{s}\right) t/4}[\gamma _{s}\cos (\gamma _{s}t/4)+(\gamma
_{s}+\gamma M)\sin (\gamma _{s}t/4)]\}/\gamma _{s}$ and $\nu =\tilde{\gamma}%
e^{\left( -4\tilde{\gamma}+\gamma _{s}\right) t/4}[\cos (\gamma
_{s}t/4)-\sin (\gamma _{s}t/4)]/\gamma _{s}$, where $\gamma _{s}=\tilde{%
\gamma}+2\gamma M$.

The time dependence of the IEP, and its geometrical bounds are explored by
numerically calculating the quantities (equations (\ref{entrop2}) and (\ref%
{lower}), (\ref{upper})) and plotted in figure 2. It is clear from figure 2
that IEP and its bounds (UB and LB) increase monotonically with time toward
the corresponding equilibrium values for a squeezed thermal bath. For the
non-squeezing case ($s=0$), our results based on quantum master equation are
fully consistent with that for the thermal bath in reference \cite{fun1}
using the method of Kraus operators. In the case of squeezed thermal bath ($%
s>0$), with increase in degree of squeezing the UB becomes higher in the
early stage of\textbf{\ }evolution $t\lesssim 0.15$, and then becomes
reduced with the growth of $s$ when $t\gtrsim 0.45$, compared with the
thermal bath ($s=0$), as shown in figure 2(a). Also, the dynamical behavior
of LB shares similar features with that of UB. Another common feature
between UB and LB is that their equilibrium times decrease with the growing
degree of squeezing, for instance, the equilibrium times for UB (LB) are 4.5
(7.5), 2 (1.4) and 0.15 (0.2) when $s=0$, $1$, and $2$, respectively. We
conclude that the summarized overall trends of geometrical bounds (LB and
UB) found in reference \cite{fun1} also hold for squeezed thermal bath,
while only the UB becomes tighter in the long-time limit, and the LB
exhibits subtle tightness in the early stage of\textbf{\ }evolution,
compared with the traditional thermal bath \cite{fun1}.

Let us next examine the time dependence of the IEP, the numerical simulation
plotted in figure 2(b), and 2(c) suggests that the values of IEP are well
limited in the region between the LB and UB, and the squeezed thermal bath
has prominent influences in both the concrete values of IEP and the critical
times $T_{c}$ of IEP reaching equilibrium, where the values of $T_{c}$
become less for growing values of $s$. Another observation in figure 2(b) is
that the deviation $\Delta \delta _{L}$ (equation (\ref{deltaL}))increases
monotonically and gradually coincides with $\Delta U$ (equation (\ref{DU})),
whereas the deviation $\Delta \delta _{U}$ (equation (\ref{deltaU}))
decreases monotonically and gradually disappears to zero in the case of $s=0$%
. With the increase of squeezing strength, i.e., $s=2$, as shown in figure
2(c), the deviation $\Delta \delta _{L}$ ($\Delta \delta _{U}$) increases
(decreases) gradually to a fixed value 0.51 (0.542), and the rates of change
of deviations $\Delta \delta _{L}$ and $\Delta \delta _{U}$ become slower.
From the above results we deduce that the actual amount of IEP becomes
depart from its UB gradually and approaches its LB with the growth of $s$.

To get a clear picture of how the IEP evolves in the parameter space of $%
\{s,t\}$, we plot the evolution of IEP in figure 3. It shows that the values
of IEP are obviously dependent on the squeezing parameters of bath. Here we
provide remarks on the parameter dependence. Although the values of $s$ are
not directly related to the system-bath interaction, whereas they depend on
both correlation time and occupation number of the bath and then affect
immensely the relaxation dynamics and steady state of the relevant system
during the thermodynamic process. As a result, any change of this key
parameter will bring significant influences on the irreversibility, and the
inherent squeezing effect stemming from the bath plays a crucial role in
understanding the relative tightness of the bounds. As we have derived
analytically in the previous paragraphs, the above summarized dependence of
bounds and IEP on the squeezing parameter is reflected in the equations (3,
4, 12).

Traditionally, the IEP could be used to evaluate the performance of
thermodynamic devices, such as the availability (or exergy) that can be
extracted from a given system, and the maximal useful work, which are
usually reduced by the presence of irreversibility \cite{TM7,book1}. It
implies that\textbf{\ }the irreversibility can be restrained by controlling
the amount of IEP through adjusting the degree of squeezing effect, as shown
in figure 3(b), where the values of IEP become less for growing values of $s$%
. From the perspective of quantum bath engineering, employing a squeezed
thermal bath is a promising avenue of using squeezing effect as a quantum
resource to improve the performance thermodynamic devices \cite%
{no8,dev1,dev8}.

As we known, squeezing effect that is rooted in Heisenberg's uncertainty
principle can be defined as the reduction in the uncertainty of some
observable, at the cost of the build-up in the conjugate one \cite%
{dev3,book2,Natt}. Physically, the squeezing involved in the bath thereby
inevitably modifies the nonunitary\textbf{\ }relaxation dynamics of system
and the relevant irreversibility during thermodynamic process. Compared with
the thermal bath, the squeezed thermal bath is taken out of thermodynamics
equilibrium through the squeezing operating, with a consequence that its
excitation number changing from\textbf{\ }$N_{th}$\textbf{\ }to\textbf{\ }$%
N=N_{th}(\cosh ^{2}r+\sinh ^{2}r)+\sinh ^{2}r$ \cite{dev3,enn}, which can be
seen as an increase in its effective temperature $T_{eff}=\omega
_{h}/k_{B}\ln [1/(N_{th}^{-1}+1)]$ with a higher frequency $\omega
_{h}>\omega _{0}$. Therefore, being purely quantum mechanical fuel in
nature, squeezed thermal bath is beneficial in its own way by providing us
with more compact energy-storage and higher effective high-temperature bath
without being actually too hot \cite{TM5}. That is to say, the squeezed
thermal state has the same entropy as the Gibbs state, but increased mean
energy, which is instrumental in the suppression of irreversibility.

\section{Geometrical bounds on irreversibility in the dephasing model}

Next we focus on the dephasing model with respect to the squeezed thermal
bath, where the bath operator is simply a sum of linear couplings to the
coordinates of a continuum of harmonic oscillators described by a spectral
density function $J(\omega )$ \cite{open1,open2,open3,open4,open5,open6},
and the decay of the coherence occurs without a decay of the corresponding
populations. Now the total Hamiltonian is

\begin{eqnarray}
H &=&H_{S}+H_{B}+H_{SB}  \notag \\
&=&\omega _{0}\hat{\sigma}_{+}\hat{\sigma}_{-}+\sum\limits_{k}\omega _{k}%
\hat{b}_{k}^{\dagger }\hat{b}_{k}+\hat{\sigma}_{z}\sum\limits_{k}(g_{k}\hat{b%
}_{k}+H.c.).  \label{H2}
\end{eqnarray}

The dynamics of the system can be characterized by the reduced density
matrix which is obtained by tracing out the degrees of freedom of the bath.
In the interaction picture, the reduced density matrix of system can be
written as \cite{open4,open5,open6,dep1,dep2}
\begin{equation}
\rho _{s}(t)=\left[
\begin{array}{cc}
\rho _{ee} & \rho _{eg}\Gamma \left( t\right) \\
\rho _{ge}\Gamma ^{\ast }\left( t\right) & \rho _{gg}%
\end{array}%
\right] ,  \label{Rec}
\end{equation}%
where the phase decay behavior of the qubit under the influence of the bath
is denoted by the factor $\Gamma \left( t\right) =Tr_{B}\rho
_{B}\prod\nolimits_{k}\exp [\alpha _{k}(t)\hat{b}_{k}^{\dagger }-\alpha
_{k}^{\ast }(t)\hat{b}_{k}]$, where $\alpha _{k}(t)=2\frac{g_{k}}{\omega _{k}%
}\left( 1-e^{i\omega _{k}t}\right) $ \cite{open4}. The associated master
equation is given by

\begin{equation}
\dot{\rho}_{s}(t)=\frac{-i\tilde{\epsilon}(t)}{2}[\sigma _{z},\rho _{s}(t)]+%
\frac{\tilde{D}(t)}{2}[\sigma _{z}\rho _{s}(t)\sigma _{z}-\rho _{s}(t)],
\label{M2}
\end{equation}%
where $\tilde{D}(t)=-\frac{d\ln (\left\vert \Gamma (t)\right\vert )}{dt}$and
$\tilde{\epsilon}(t)=-$Im$[\frac{d\Gamma (t)/dt}{\Gamma (t)}].$

In the following we consider that the bath starts from a squeezed thermal
state \cite{dep1,dep2,therm1,therm2}

\begin{equation}
\rho _{B}(0)=\hat{\varsigma}\rho _{th}\hat{\varsigma}^{\dagger },
\label{rou}
\end{equation}%
where $\rho _{th}=e^{-\beta H_{B}}/Z_{\beta }$ is thermal state with $%
Z_{\beta }$ the partition function. $\hat{\varsigma}$ $=\sum\nolimits_{k}%
\hat{s}_{k}$, where $\hat{s}_{k}=e^{[(s_{k}^{\ast }e^{-i\phi _{k}}\hat{b}%
_{k}^{2}-s_{k}e^{i\phi _{k}}(\hat{b}_{k}^{\dagger })^{2})/2]}$ is the
squeezing operator for the boson bath mode $\hat{b}_{k}$ with $s_{k}$ and $%
\phi _{k}$ being the bath squeezing strength and phase, respectively. In
this situation, the function $\Gamma \left( t\right) $ could be evaluated
under the summation of the modes of the squeezed thermal bath as \cite%
{open4,open5,open6,dep1,dep2}
\begin{equation}
\Gamma \left( t\right) =e^{[-\sum_{k}\frac{4\left\vert g_{k}\right\vert ^{2}%
}{\omega _{k}^{2}}\left( 1-\cos \omega _{k}t\right) \gamma _{k}\left(
t\right) \coth (\frac{\omega _{k}}{2T})]}  \label{dep1}
\end{equation}%
with $\gamma _{k}\left( t\right) =\cosh 2s_{k}-\sinh 2s_{k}\cos \left(
\omega _{k}t-\Delta \phi _{k}\right) $ and $\Delta \phi _{k}$ is the phase
difference between the squeezing phase $\phi _{k}$\ relative to the phase of
the coupling strength $g_{k}$.

\begin{figure}[!ht]
\centering
\includegraphics[width=12cm]{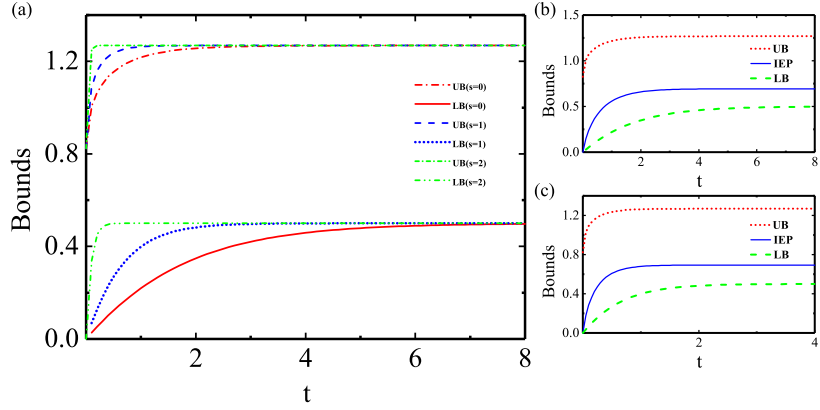}
\caption{(a) Time dependence of LB and UB of IEP under different degrees of
squeezing $s$ in the dephasing model. Time dependence of IEP (blue solid
line) and its LB (red dotted line) and UB (green dashed line) in the case of
(b) $s=0$ and (c) $s=1$. The initial state of system is $\left( \left\vert
e\right\rangle +\left\vert g\right\rangle \right) /\protect\sqrt{2}$, and
the parameter $\Delta \protect\phi =\protect\pi /4$.}
\end{figure}

Substituting the coupling spectral density $J(\omega )=2\pi
\sum_{k}\left\vert g_{k}\right\vert ^{2}\delta \left( \omega -\omega
_{k}\right) $ into the equation (\ref{dep1}), we can transform the above
summation in $\Gamma \left( t\right) $ into an integral for continuous bath
modes as

\begin{eqnarray}
\Gamma \left( t\right) &=&\exp \{-\int_{0}^{\infty }\frac{d\omega }{\pi
\omega ^{2}}2J(\omega )(1-\cos \omega t)\coth (\omega /2T)  \notag \\
&&\times \left[ \cosh (2s)-\sinh (2s)\cos \left( \omega t-\Delta \phi
\right) \right] \}.  \label{dep2}
\end{eqnarray}%
\begin{figure*}[tbp]
\centering\includegraphics[width=14cm,bb=3pt 14pt 872pt 390pt]{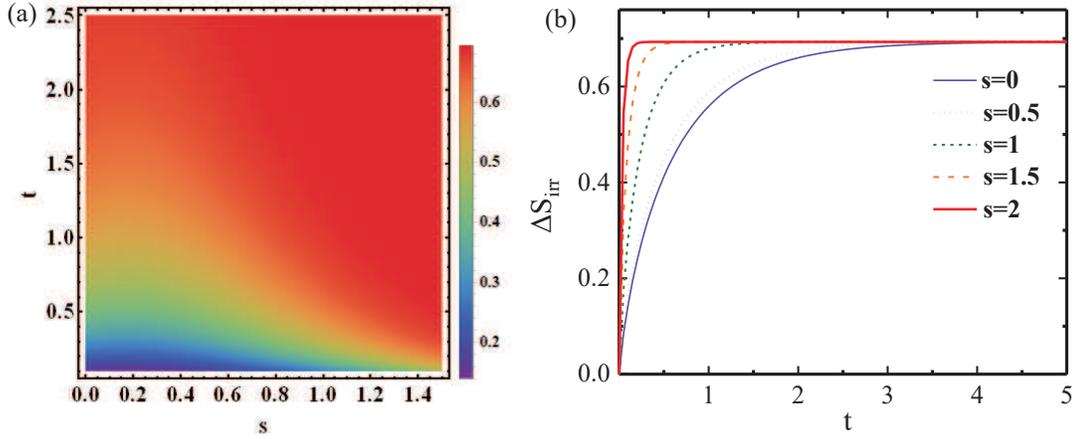}
\caption{(a) The values of the IEP $\triangle S_{irr}$ in the parameter
plane of $\{s,t\}$ in the dephasing model. (b) Time dependence of IEP $%
\triangle S_{irr}$ under the different degrees of squeezing $s$. Here the
parameter $\Delta \protect\phi =\protect\pi /4$.}
\end{figure*}
In the present work we adopt the Ohmic coupling spectral density $J(\omega
)=\eta \omega e^{-\omega /\varpi }$ with $\varpi $ cutoff frequency and $%
\eta $ is the coupling strength \cite{noise}. Note that such engineering of
the spectrum's Ohmicity seems possible when simulating the dephasing model
using trapped ultracold atom, as demonstrated in reference \cite{atom1,atom2}%
. In the high-temperature regime, the dephasing process is Markovian, and
after straightforward algebra, one finds
\begin{equation}
\Gamma \left( t\right) =e^{\{-\frac{2\eta Tt}{\pi }\left[ \pi \cosh (2s)-\ln
4\sinh (2s)\sin \Delta \phi \right] \}},  \label{dep3}
\end{equation}%
where the approximation $\coth (\omega /2T)\approx 2T/\omega $ has been
taken. Expression (equation (\ref{dep3})) is the exact analytic for the
time-dependent dephasing rate $\Gamma \left( t\right) $ in the present model.

Special attention was paid to the time dependence of the IEP as well as its
geometrical bounds, as shown in figure 4, where one can find that all the
quantities exhibit asymptotic behaviors approaching their stationary values.
They correspond to relaxations of the system through the dephasing channel
due to the system-bath coupling. Regarding the time dependence of the
bounds, we encounter another common feature is that the equilibrium times of
bounds decrease with the growing degree of squeezing, for instance, the
equilibrium times for UB (LB) are 3 (7.3), 2.5 (3.3) and 0.4 (0.6) when $s=0$%
, $1$, and $2$, respectively. But unlike the dissipation model, both UB and
LB increase monotonically as the squeezing character of the bath grows, and
only the LB becomes tighter (compared with the thermal bath) during the
whole dynamic process in the dephasing model. As a contrast, the tightness
of LB only appears in the early stage of evolution for the dissipation
model. By comparing the time evolutions of IEP with the two blue solid-lines
in panel (b,c) of figure 4, we find that the IEP is well located inside the
region between the LB and UB, and reaches its stationary value faster with
the increase of the squeezing parameter $s$, and the values of the IEP is
reduced due to the existence of squeezing effect.

In figure 5, we provide numerical estimates of the IEP in the parameter
space of $\{s,t\}$. The figure 5 tells us that one can precisely control the
thermodynamic irreversibility through adjusting the parameters of the bath.
As shown in figure 5(b), under the dephasing model, the IPE reaches
equilibrium faster as the squeezed parameter increases, and the value of the
IEP in long-time limit $\bigtriangleup S_{irr}(\infty )$ will eventually
converge together, irrespective of the values of parameter $s$. It means
that, in the dephasing model, the squeezing effect could not make too much
impact on the thermodynamic irreversibility in the long-time limit, although
the existence of squeezing drives the system into equilibrium faster.
Physically, on a fundamental level, quantum coherence and the related
dephasing process could also alter the possible state transitions in
thermodynamic processes \cite{Los}, and may even modify the
fluctuation--dissipation relation \cite{FD1,FD2} and quantum nonequilibrium
work relation \cite{smith}. Additionally, when a system relaxes to
equilibrium through contacting with a thermal bath, quantum coherences are
known to contribute an additional term to the IEP \cite{Iep1,Iep2,Iep3}.
Different from the dissipation case where the system can exchange energy
with its bath, in the dephasing model, this open system can never exchange
energy with its bath. But the information and correlation exchange between
the system and the bath are dominant during the dynamics and this exchange
also influence the IEP. As a result, any alteration in the von Neumann
entropy (basis of IEP and its bound) resulting from the relaxation process
(dissipation or dephasing) has contributions not only from the change in
population but also from decoherence. In this regard, it was pointed out
that the entropy production can be split in two contributions, an incoherent
one (stemming from populations) and a coherent one (stemming from quantum
coherences) \cite{ep1,ep2,ep3}.

\section{Discussion and conclusions}

The study of IEP is of importance due to its intimate relation with the
arrow of time in classical and quantum systems \cite{arr1,arr2}, the SLT
\cite{law1,law2,law3},\textbf{\ }thermodynamic operations and thermal
machines \cite{mac1,mac2,mac3,mac4}, and quantum and classical speed limits
\cite{TM8,speed1,speed2}. Therfore,\ tightening the bounds of IEP not only
deepens our understanding of how much entropy production changes during the
thermodynamic process but also provides insights into how to improve the
performance of\textbf{\ }quantum thermodynamic devices.

Interaction with a squeezed thermal bath is not the only generalized process
that goes beyond the typical settings in classical thermodynamics.\textbf{\ }%
Our findings demonstrate how to utilize the squeezing effect of bath as a
resource to control the irreversibility, where the use of nonthermal bath
offers more degrees of control and manipulation, such as the amount of
squeezing. Note that quantum bath engineering techniques has become powerful
tools that enable the realization of arbitrary thermal and nonthermal bath,
for instance,\emph{\ }experimental realizations of squeezed thermal states
range from superconducting circuit QED \cite{mw1,mw2,mw3} to optomechanical
mechanical oscillators \cite{osc1,osc2}. The key parameters considered in
our numerical simulation, such as the inverse temperature $\beta $ and the
degree of squeezing $s$, could be experimentally controlled using the
current technologies demonstrated in the above-mentioned experiments.
Additionally, there have been many experiments focus on the assessment of
nonequilibrium thermodynamics irreversibility using the technology of
quantum trajectories of stochastic dynamics in nuclear magnetic resonance
setups \cite{TM3}, superconducting qubit \cite{sc}, and mechanical resonator
\cite{MR}, respectively. Our results reveal more detailed properties of
thermodynamic irreversibility that are stronger than the conventional SLT,
for\textbf{\ }given a restricted class of irreversible processes. Along with
other studies addressing squeezing effects in quantum thermodynamics, we
hope that our analyses help to unveil the role of squeezing effects in
quantum thermodynamics devices.

In summary,\textbf{\ }we studied the influence of squeezed characteristics
of bath on the IEP of open quantum systems. The results show that the
equilibrium rates of IEP and its bounds become faster, and the values of IEP
are reduced through harvesting the benefits of squeezing effects in the case
of both dissipation model and dephasing model. In the dissipation model, the
summarized overall trends of geometrical bounds (LB and UB) found in
reference \cite{fun1} also hold for squeezed thermal bath, while only the UB
becomes tighter in the long-time limit, and the LB exhibits subtle tightness
in the early stage of\textbf{\ }evolution, compared with the thermal bath.
Unlike the dissipation model, both UB and LB increase monotonically as the
squeezing character of the bath grows, and only the LB becomes tighter
(compared with the thermal bath) during the whole dynamic process in the
dephasing model.

Moreover, the concrete amount of thermodynamic bounds greatly depends on the
explicit form of the system-bath coupling, whereas the trends of them is
independent of the details of the models. Also, the above-summarized trends
for the bounds are independent of system size and hold for systems having
more degrees of freedom. Our results do not contradict the SLT, which is
modified by the inclusion of squeezing as an available resource in the bath.
It is worth noting that a general evolution and the associated geometrical
bounds of irreversibility of two-level system in thermal bath were
theoretically analyzed and experimentally demonstrated in \cite{fun1}. Here,
we further highlight the role of adjustable parameters in bath, such as
temperture and squeezing degree, on the reduction of thermodynamic
irreversibility. It is expected that the present work helps in developing a
better understanding of the irreversibility under ambient conditions.

\section*{ACKNOWLEDGMENT}

The authors thank Jun-Hong An and Zhang-Qi Yin enlightening discussions.
This work is supported by the Hubei Province Science Fund for Distinguished
Young Scholars under Grant No. 2020CFA078, and by the Special Project for
Research and Development in Key areas of Guangdong Province under Grant No.
2020B0303300001, National Natural Science Foundation of China under Grants
No. U21A20434, No. 12074390, No. 11835011, No. 11734018. JBY acknowledges supports from A*STAR Career Development Award (C210112010), A*STAR (\#21709), and National Research Foundation Singapore via Grant No. NRF2021-QEP2-02-P01.

\end{document}